\documentclass[12pt,preprint]{aastex}\usepackage{amssymb,psfig}
\newcommand{\beq}{\begin{equation}}
\newcommand{\eeq}{\end{equation}}
\newcommand{\bea}{\begin{eqnarray}}
\newcommand{\eea}{\end{eqnarray}}

\newcommand{\rem}[1]{ }
\bibpunct[,]{(}{)}{;}{a}{}{,}
\begin{document}

\title{Boundary Layer Self-Similar Solution for the Hot Radiative Accretion 
onto a rapidly Spinning Neutron Star}

\author{Mikhail V. Medvedev 
\thanks{
Also at the Institute for Nuclear Fusion, RRC ``Kurchatov
Institute'', Moscow 123182, Russia}
\\
Department of Physics and Astronomy, \\
University of Kansas, KS 66045}


\begin{abstract}
We consider hot accretion onto a rapidly spinning neutron star
(or any other compact object with a surface). A radiative 
hot settling flow has been discovered at low accretion rates
in the early work by \citet{MN01} and analytical solution 
has been presented. It was shown later that this flow can
match external medium smoothly, thus enforcing its physical 
feasibility. Here we complete the study of the global structure
of such hot accretion by presenting the analytical solution for 
the boundary later, which forms between the bulk of the flow and
the stellar surface. We confirm our results via full numerical 
solution of height-integrated two-temperature hydrodynamic equations.
\end{abstract}

\section{Introduction}

Accretion flows around compact objects frequently radiate significant
levels of hard X-rays, indicating the presence of hot optically-thin
gas in these systems.  This has motivated the study of hot accretion
flows around compact stars.

Various models of accretion may be divided into two large groups
according to whether they require a shock to form near the central 
object (e.g., a neutron star) or not. The first group includes
spherical accretion flows onto a neutron star (NS) and/or a black hole 
(BH) in either a kinetic \citep{ZS69,AW73,Tur94,Zam95,Zan98} or a fluid
(hydrodynamic) regime \citep{SS75,CS97,KL82}. Whereas the existence
of a shock is not a problem in spherical flows, the shock discontinuity
in rotating flows likely leads to serious physical inconsistencies
associated with causality of viscous processes \citep{P77,PN92}
[except for the case of ``gap accretion'' in which the surface of the 
accreting star lies below the marginally stable orbit and, hence, is
not in causal contact with the bulk of the flow \citep{KW91,D01}].

The other group includes flows which contain no shocks, which
(or, at least, the inner regions of which) thus, must be subsonic and 
have a gas heated up to the nearly virial temperature. Such hot boundary
layers form in the cool thin disk accretion, provided a causal
viscosity prescription is used \citep{PN92,NP93,PS01}. A large class of 
hot accretion flows: the SLE solution \citep{SLE76}, the
advection-dominated accretion flow (ADAF)
\citep{I77,R82,NY94,NY95a,NY95b,Abram+95}, the advection-dominated
inflow-outflow solution (ADIOS) \citep{BB99}, and the
convection-dominated accretion flow (CDAF) \citep{NIA99,QG99}. 
The relevance of these solutions for accretion onto a NS is unclear.
Moreover, since the marginally stable orbit appears not to play an important
role in hot quasi-spherical flows \citep{N97,C97}, it is, therefore, not 
clear that one would necessarily have a shock with a hot flow even if the
accreting star were very compact. 
Finally, a subsonic hot accretion flow may form
around a magnetized neutron star in the propeller state \citep{DP81,I01,I03}
[in which the gas heating by viscosity is much faster than radiative cooling]
and the ``hot settling flow'' \citep{MN01,NM03} [in which heating and 
cooling balance each other] definitely fall in this group.

The latter, hot settling flow,  could equally well be described as a 
``hot atmosphere'' since the solution is, to first approximation, 
static, and accretion represents only a small perturbation on the static 
solution.  To our knowledge, the hot settling flow is the only steady state 
solution for accretion onto a NS presently available that does not involve 
a discontinuity near the surface of the star. The hot settling flow exists 
at rather low accretion rates, $\dot M$, smaller than a few percent of 
Eddington. The flow is subsonic everywhere; it is powered by
the rotational energy of the central accretor which is braked by
viscous torques. A very interesting property of the flow is that,
except for the inflow velocity, all gas properties, such as density,
temperature, angular velocity, luminosity, and angular momentum flux,
are independent of the mass accretion rate; the flow properties are 
sensitive to the star spin, $s$ \citep{MN01}. Because too few parameters
control the flow structure (effectively, just $\dot M$ and $s$) it was 
not clear whether the flow can match to a general external medium
(for a fixed spin of a star). Recently, \citet{NM03} demonstrated that
the flow matches the external medium via a special ``bridging solution'',
which forms in a relatively narrow region near the outer boundary.

To complete the study of the structure of hot settling accretion flow
onto a neutron star, it is, thus, required to study how this flow
matches the inner boundary conditions set by the parameters of the star.
The study of the inner boundary layer is presented in this paper.
We present an analytical self-similar solution and confirm it numerically.
Our results are in agreement with other studies \citep{TLM98,TO99}, and 
may be important for the interpretation of kHz quasi-periodic oscillations. 
We critically examine the limitations of our model as well.

\section{The model}

We consider viscous hydrodynamic accretion onto a compact spinning object
with a surface. The central object has a radius $R_*=r_* R_g$ 
(where $R_g=2GM_*/c^2=2.8\times10^5 m$~ cm is the Schwarzchild radius), a mass
$M_*=m M_{\sun}$, and an angular velocity $\Omega_*=s \Omega_{K*}$,
where $\Omega_K(R)=(GM_*/R^3)^{1/2}$ is the Keplerian angular velocity
at radius $R$ and $\Omega_{K*}=\Omega_K(R_*)$.  The mass accretion rate
is $\dot M=\dot m\dot M_{\rm Edd}$, where 
$\dot M_{\rm Edd}=1.39\times10^{18}m$~g~s$^{-1}$ (corresponding to a radiative
efficiency of 10\%). We use the height-integrated form of the viscous 
hydrodynamic equations \citep{I77,A+88,Paczynski91,NY94}.

In standard accretion problems the radial coordinate, $R$, varies through 
a large range. Hence an analytical self-similar solution, in which the gas
parameters (density, temperature, and such) scale as power-laws of $R$, 
is usually possible in the region far from the boundaries. 

Unlike the bulk of the flow, the solution to the boundary layer (BL) cannot
be obtained in a self-similar form in terms of the radial coordinate $R$.
Indeed, the structure of the BL is intrinsically non-self-similar in $R$ 
because all the gas parameters (e.g., the temperature, gas density, etc.) 
change very dramatically over a relatively short radial region: 
$R_*\le R \la 2 R_*$. For instance, the density nearly diverges as one 
gets close to the star surface whereas the temperature
decreases to the values well below the virial temperature.  
Such a behavior, however, suggests to look for a self-similar solution
in terms of the distance from the stellar surface, i.e., in terms of 
\beq
D=R-R_*.
\eeq 
In our calculation we neglect the effects of radiation
transfer and Comptonization.
They may be important in hot regions, but will unlikely strongly
affect the flow closer to the star, where the temperature of the
gas falls below few$\times10^{9}$~K or so (see discussion in 
Section \ref{s:discuss}). 

Unlike the hot settling flow case, here we cannot neglect the radial (infall)
velocity. We use the height-integrated two-temperature hydrodynamic 
equations, written in the approximation that $R=R_*+D$ with $D\ll R_*$~:
\bea
& \displaystyle -\dot M=4\pi R_* \rho v, & 
\label{1}\\
& \displaystyle v\frac{dv}{dD}=\left(\Omega^2-\Omega_{K*}^2\right)R_* -
\frac{1}{\rho}\,\frac{d}{dD}\left(\rho c_s^2\right), & 
\label{2}\\
& \displaystyle 4\pi\alpha\left(\rho c_s^2\right)\frac{R_*^4}{\Omega_{K*}}\,
\frac{d\Omega}{dD}=\dot J-\dot M\Omega R_*^2, & 
\label{3}\\
& \displaystyle \frac{\rho v}{\gamma_p-1}\,\frac{d c_{sp}^2}{dD}
-c_{sp}^2 v\frac{d\rho}{dD}=q^+-q_{\rm Coul}, & 
\label{4}\\
& \displaystyle \frac{\rho_e v}{\gamma_e-1}\,\frac{d c_{se}^2}{dD}
-c_{se}^2 v\frac{d\rho_e}{dD}=q_{\rm Coul}-q^-, & 
\label{5}
\eea
where $\rho=\rho_p+\rho_e\approx\rho_p$ is the mass density of the 
accreting gas ($\rho_p$ and $\rho_e\simeq(m_e/m_p)\rho$ are the mass 
densities of the proton and electron fluid), $v$ is the
radial infall velocity [note that in equation (\ref{1}) we took 
into account that the radial velocity is negative (inward)], 
$\Omega$ is the angular velocity, $c_s^2=c_{sp}^2+c_{se}^2\, m_e/m_p$ 
is the square of the isothermal non-relativistic sound speed 
[this follows from the fact 
that both electrons and protons contribute to the total gas pressure:
$\rho c_s^2=p=p_p+p_e=\rho_p c_{sp}+\rho_e c_{se}^2
\approx\rho(c_{sp}^2+c_{se}^2\, m_e/m_p)$] and $c_{sp}$ and $c_{se}$
are the respective isothermal sound speeds of the two species,
$\alpha$ is the Shakura-Sunyaev viscosity parameter,
$\dot J$=const. is the rate of accretion of angular momentum 
(an eigenvalue of a problem),
$\gamma_p$ and $\gamma_e$ are the adiabatic indexes of
protons and electrons, which we assume to be equal, for simplicity.
We assumed that the flow is hot and quasi-spherical, i.e., the local 
vertical scale height $H=c_s/\Omega_K$ is comparable to the local radius 
$R\simeq R_*$.  We may then set, for simplicity, that $H=R$ \citep{MN01}.
Finally, $q^+$, $q^-$, and $q_{\rm Coul}$ are the viscous
heating rate, radiative cooling rate, and energy transfer rate from
protons to electrons via Coulomb collisions, per unit mass. 
We have assumed that all viscous heat goes into the proton component.
We also assume that the gas is optically thin.

The temperature of the gas determines the efficiency of the Coulomb 
energy transfer from the protons to the electrons and the rate of 
Bremsstrahlung cooling of the electrons. The balance between them
defines whether the gas is in the two temperature regime (when the 
temperatures are high and the Coulomb collisions are very rare) or
in the one-temperature regime (when temperatures are lower). 

In the two-temperature regime, we expect that 
the electrons are relativistic and the protons are non-relativistic:
$c_{se}^2/c^2=kT_e/(m_ec^2)\gg 1,\ c_{sp}^2/c^2=kT_p/(m_pc^2)\ll1$.  
The viscous heating rate of the gas,
the energy transfer rate from the protons to the electrons via Coulomb
collisions, and the cooling rate of the electrons via Bremsstrahlung
emission are given by
\bea 
q^+&=&\alpha\left(\rho c_s^2\right)\,\frac{R_*^2}{\Omega_{K*}}
\left(\frac{d\Omega}{dD}\right)^2,
\label{q+}\\
q_{\rm Coul}&=&Q_{\rm Coul}\,\rho^2\frac{c_{sp}^2}{c_{se}^2} ,\quad
Q_{\rm Coul}=4\pi r_e^2 \ln{\Lambda} \frac{m_ec^3}{m_p^2}, 
\label{qc}\\
q^-&=&Q_{\rm ff,R}\,\rho^2 c_{se}^2 ,\quad
Q_{\rm ff,R}=48\alpha_f r_e^2\frac{m_ec}{m_p^2} ,
\label{q-R}
\eea 
where $\alpha_f$ is the fine structure constant, $r_e$ is the
classical electron radius, $\ln{\Lambda}\simeq20$ is the Coulomb
logarithm, $c_s^2\simeq c_{sp}^2$, and we have neglected logarithmic
corrections to the relativistic free-free emissivity. The subscript
``R'' in $Q_{\rm ff,R}$ denotes relativistic Bremsstrahlung.

In the one-temperature regime, both protons and electrons are cool 
and non-relativistic, and have nearly the same temperature, hence
$c_{sp}^2\simeq(m_e/m_p)c_{se}^2$ and $c_s^2\approx 2c_{sp}^2$.  
The two energy equations (\ref{4}) and (\ref{5}) can be combined 
to yield a single energy equation for the accreting gas:
\beq
\displaystyle \frac{\rho v}{\gamma-1}\,\frac{d c_s^2}{dD}
-c_s^2 v\frac{d\rho}{dD}=q^+-q^-, 
\label{45}
\eeq
where the free-free cooling takes the form
\beq
q^-= Q_{\rm ff,NR}\rho^2 c_{s}, \quad 
Q_{\rm ff,NR}=5\pi^{-3/2}\alpha_f\sigma_T\frac{m_e^{1/2}c^2}{m_p^{3/2}},
\eeq
where $\sigma_T$ is the Thompson cross-section, and the subscript
${\rm NR}$ stands for non-relativistic. 

As a result of high density of the gas in the BL, optically thin 
bremsstrahlung cooling dominates over self-absorbed synchrotron cooling;
hence the latter may safely be neglected. We therefore neglect synchrotron 
emission in our analysis. In our model, we neglect the effects of radiation 
pressure compared to the gas pressure (which is of order 
$\dot M/\dot M_{\rm Edd}\ll1$ and, hence, negligible at low accretion rates)
and Comptonization (which must be important in a high-temperature region,
but is not important closer to the stellar surface, where the gas 
temperature is low, see more discussion below). 
For simplicity, we neglect also thermal conduction. 
Thus, our present model is similar to the models we used in our
previous studies of hot accretion. Our simplified hydrodynamic model
is, therefore, very instructive. It allows us to study the BL problem
on the same grounds, on which the other hot flow solutions have been treated,
--- as a viscous, radiative, purely hydrodynamic flow. 
Once the basic hydrodynamic structure of the BL is understood, it will be
worthwhile to put forward more sophisticated and detailed models, which should
involve additional physics neglected in the present analysis. 

The set of equations (\ref{1})--(\ref{5}) must satisfy certain 
boundary conditions at the neutron star.  As the flow approaches the
surface of the star, the radial velocity must become
very much smaller than the local free-fall velocity. For the flow
with $v\not=0$ to match the radially non-moving stellar surface, the 
gas density should diverge, according to Eq. (\ref{1}). Being proportional 
to the density squared, radiative cooling diverges as well, thus bringing 
the gas temperature to zero. In contrast, the angular velocity must approach 
the angular velocity of the star $\Omega_*$, hence remains nonzero.  
Naturally, the inner boundary conditions, as $D\equiv (R-R_*)\to0$,  
are\footnote{These conditions correspond to an infinitely dense and cold 
star. To allow for a smooth match of a BL solution with a realistic,
non-zero temperature star, an additional physics, such as thermal
conduction, must be invoked. Such a consideration goes beyond the 
scope of the present study.}
\beq
\rho\to\infty,\qquad c_s^2\to0,\qquad \Omega\to\Omega_*,\qquad v\to0.
\eeq
The outer boundary conditions are set by the hot settling flow solution
at $R\to R_*$ \citep{MN01}.
We treat the mass accretion rate $\dot M$ as a parameter.

We would like to comment here that a self-similar solution is only an
apptoximate solution far from the boundaries.  In order to match boundary
conditions, transition layers can form where variables relaxe
from the values at the boundary to a self-similar behavior.
These layers may be quite extended and be the site of interesting
physical processes not accurately reproduced by the self-similar
solution. As we shall see below, the self-similar solution for the 
hot boundary layer matches well the conditions at the inner 
boundary, whereas at the outer boundary the flow is required to settle
onto a subsonic, hot inflow and a transition layer appears, see Figure
\ref{f:BL} in the next Section.

\section{Self-similar solution for a boundary layer}  
\subsection{Two-temperature solution}

The gas in the two-temperature regime is governed by equations
(\ref{1})--(\ref{5}), which we now consider one by one and identify
leading terms in them.

Let us first consider equation (\ref{2}). First, we note that the rotation
of the gas is sub-Keplerian, $\Omega_*^2\ll \Omega_{K*}^2$, 
so that we can neglect the first term in equation (\ref{3}). 
Next, we cast the equation into the form:
\beq
\frac{d}{dD}\left(c_s^2+\frac{1}{2}v^2 \right)+
c_s^2\left(\frac{1}{\rho}\,\frac{d\rho}{dD}
+\frac{1}{2}\,\frac{v_{\rm ff,*}}{c_s^2 R_*} \right)=0,
\label{momnt}
\eeq
where $v_{\rm ff,*}=\sqrt{2}\Omega_{k*}R_*$ is the free-fall velocity 
that near the stellar surface. We now make the following assumptions,
which consistency with the obtained solution must be checked
{\it a posteriori}: (i) the flow is always subsonic, $v^2\ll c_s^2$, then the
second term in the first brackets may be neglected and
(ii) $c_s^2$ grows with $D$ slower than linearly 
(for $c_s^2\propto D^{\beta}$ we should have $\beta<1$), then
the second term in the second brackets is sub-dominant and may be 
neglected as well. With these assumptions, the equation simplifies to
\beq
\rho c_s^2 = p = \textrm{ const.},
\label{2r}
\eeq
that is, the pressure is constant throughout the boundary layer.
Note also that in the two-temperature regime, the gas pressure is
dominated by the protons: 
\beq
\rho c_s^2
=\rho\,\left(\frac{kT_p}{m_p}+\frac{kT_e}{m_e}\,\frac{m_e}{m_p}\right)
\approx\rho\,\frac{kT_p}{m_p}= \rho c_{sp}^2,
\eeq
because $T_p\gg T_e$, where $T_p$ and $T_e$ are the proton and
electron temperatures.

As we mentioned earlier, we are looking for the solution which is self-similar 
(i.e., power-law) in $D$. On the other hand, angular velocity $\Omega$
approaches a constant at the star surface: $\Omega=\Omega_*$. 
Thus, we readily conclude that $\Omega\propto D^0$ (otherwise it is 
either zero or diverges at $D=0$). Therefore, equation (\ref{3})
ought to reduce to
\beq
d\Omega/dD=0
\label{3r}
\eeq
with the right-hand-side of (\ref{3}) being sub-dominant (i.e., it 
will introduce a small $D$-dependent correction to the zeroth order
solution $\Omega=\Omega_*$; we demonstrate this below).

Since $d\Omega/dD=0$, the heating rate in equation (\ref{4}) vanishes. 
Together with the continuity equation (\ref{1}), the energy equations 
for the protons and the electrons read,
\bea
& &\displaystyle{ \frac{\dot M}{4\pi R_*^2}\left(\frac{1}{\gamma-1}\,
\frac{d\,c_{sp}^2}{dD}-\frac{c_{sp}^2}{\rho}\,\frac{d\rho}{dD}\right)
=Q_{\rm Coul}\,\rho^2\frac{c_{sp}^2}{c_{se}^2}, }
\label{4r}\\
& &\displaystyle{ \frac{\dot M}{4\pi R_*^2}\,\frac{m_e}{m_p}\,
\left(\frac{1}{\gamma-1}\,
\frac{d\,c_{se}^2}{dD}-\frac{c_{se}^2}{\rho}\,\frac{d\rho}{dD}\right)
=Q_{\rm ff,R}\rho^2 c_{se}^2-Q_{\rm Coul}\,\rho^2\frac{c_{sp}^2}{c_{se}^2}, }
\label{5r}
\eea

The system of equations (\ref{1}), (\ref{2r}), (\ref{3r})--(\ref{5r})
 admits the following self-similar solution:
\beq
\rho=\rho_0\, d^{-2/5},\quad c_{sp}^2=c_{sp0}^2\, d^{2/5},
\quad c_{se}^2=c_{se0}^2\, d^{1/5},
\quad v=v_0\, d^{2/5},\quad \Omega=\Omega_0\, d^0,
\label{sol2T}
\eeq
where we used the dimensionless distance $d=D/R_*$. The constant factors may 
be found as follows. Equation (\ref{4r}) determines the relation between
$c_{se0}^2$ and $\rho_0$. The sum of equations (\ref{4r}) and (\ref{5r}),
together with $c_{se}^2(m_e/m_p)\ll c_{sp}^2$ and Eqs. (\ref{sol2T}), 
yields the relation between $c_{s0}^2$ and $\rho_0$. On the other hand, 
at the outer edge, BL should smoothly match to the hot settling solution.
This requires the pressure balance to hold at the interface,
hence $p=p_0=\rho_0 c_{s0}^2=p_{\rm out}$. From this condition, all the
constants are determined uniquely (the angular velocity is, of course,
set by the stellar rotation):
\bea
& &\displaystyle{\rho_0=p_{\rm out}^{1/5}
\left(Q_{\rm ff,R}Q_{\rm Coul}/c^2\right)^{-1/5} B^{2/5}, }\\
& &\displaystyle{c_{sp0}^2=\rho_0^4 
\left(Q_{\rm ff,R}Q_{\rm Coul}/c^2\right) B^{-2}, }\\
& &\displaystyle{c_{se0}^2=\rho_0^2 Q_{\rm Coul} B^{-1}, }\\
& &\displaystyle{v_0={\dot M}/\left({4\pi R_*^2\rho_0}\right), } 
\label{v1}\\
& &\displaystyle{\Omega_0=\Omega_*,}
\label{Omega1}
\eea
where we introduced
\beq
B=\frac{\dot M}{10\pi R_*^3}\,\left(\frac{\gamma}{\gamma-1}\right).
\eeq
Finally, $p_{\rm out}$ is estimated from the self-similar settling 
flow solution \citep{MN01} by setting the transition radius approximately 
equal to the radius of the star: 
\beq
p_{\rm out}=\frac{\sqrt{3}}{16}\,
\frac{c^4}{\sqrt{Q_{\rm ff,R}Q_{\rm Coul}} R_g}\,
\left(\alpha s^2 r_*^{-3}\right)\,.
\label{pout}
\eeq

The transition from the boundary layer to the hot settling flow
occurs at some distance $d_{tr}$, which can be estimated by matching
the boundary layer density or sound speed (which is equivalent,
because the pressure also matches) to that of the hot settling flow.
Assuming that $d_{tr}\ll 1$, the hot settling flow has the proton 
sound speed squared is equal to $(c^2/6)r_*^{-1}$. Therefore, from 
$c_{s0}^2\sim (c^2/6)r_*^{-1}$, we obtain:
\beq
d_{\rm tr}\simeq(6 r_*c_{sp0}^2/c^2)^{-5/2}\sim 0.044
\label{d-tr}
\eeq
for typical parameters, $\alpha=0.1$, $\dot m=0.01$, $s=0.3$, $m=1.4$, 
$r_*=3$.

\subsection{One-temperature solution}

In the one-temperature regime, the temperatures of the protons and 
electrons are nearly equal, $c_{sp}^2\approx (m_e/m_p)c_{se}^2$, and the
both species contribute equally to the pressure. 
The energy equations (\ref{4r}) and (\ref{5r})
reduce to the single energy equation for the accreting gas (\ref{45}).
Since $d\Omega/dD=0$, the heating rate in equation (\ref{45}) vanishes. 
Together with the continuity equation (\ref{1}), the energy equation reads,
\beq
\frac{\dot M}{4\pi R_*^2}\left(\frac{1}{\gamma-1}\,\frac{d\,c_s^2}{dD}
-\frac{c_s^2}{\rho}\,\frac{d\rho}{dD}\right)=Q_{\rm ff,NR}\rho^2 c_s
\label{45r}
\eeq

The system of equations (\ref{1}), (\ref{2r}), (\ref{3r}), (\ref{45r}), 
admits the following self-similar solution:
\beq
\rho=\rho_0\, d^{-2/5},\quad c_s^2=c_{s0}^2\, d^{2/5},
\quad v=v_0\, d^{2/5},\quad \Omega=\Omega_0\, d^0.
\label{sol1T}
\eeq
Here the constant factors are:
\bea
& &\displaystyle{\rho_0=p_{\rm out}^{1/5} Q_{\rm ff,NR}^{-2/5} B^{2/5}, }\\
& &\displaystyle{c_{s0}^2=\rho_0^4 Q_{\rm ff,R}^2 B^{-2}, }
\eea
and $v_0$ and $\Omega_0$ are given by equations (\ref{v1}), (\ref{Omega1}).

\subsection{Some comments}

Having derived the self-similar solutions, we now prove   
that the assumptions made in order to simplify equations 
(\ref{2}) and (\ref{3}) are consistent.

Let us first consider the angular momentum equation (\ref{3}). 
Because $\rho c_s^2=p_{\rm out}=$const. in the boundary layer
and angular momentum flux $\dot J=$const. throughout the entire 
accretion flow, this equation has a simple solution:
\bea
\Omega&=&\Omega_*\,\left[\frac{\dot J}{\dot M\Omega_* R_*^2}\,
\left(1-e^{-D/A}\right)+e^{-D/A}\right]
\nonumber \\
&\simeq&\Omega_*\,\left[1+\left(\frac{\dot J}{\dot M\Omega_* R_*^2}-1
\right)\,
\frac{D}{A} \right]
\equiv\Omega_*\,(1+\epsilon\,d),
\eea
where 
\beq
A\equiv 4\pi\alpha\, p_{\rm out}\,{R_*^2}/{(\Omega_{K*}\dot M)}
\simeq 10^{-2}\alpha^2s^2\dot m^{-1} r_*^{1/2}\,R_g,
\eeq
and in the expansion we assumed $D/A\ll 1$. 
Note that for typical parameters of the hot settling flow:
$\alpha\sim0.1,\ s\sim0.1,\ \dot m\la0.01,\ r_*\simeq3$, we
estimate $A\ga10^{-4} R_g$. To calculate the 
first-order correction to the solution, $\epsilon$, we use the
expression for $\dot J$ from Eq. (15f) of \citet{MN01}. Noting that
$\dot J/(\dot M\Omega_* R_*^2)\gg1$, we obtain:
\beq
\epsilon\simeq-(3/2)r_*^{-3}\sim(1/18)\ll1;
\eeq
note that $\epsilon$ depends on $r_*$ only.
Therefore the zeroth order solution describes the scaling of $\Omega$
accurately throughout the boundary layer (where $D/A\la1$), with the 
first order correction introducing just a minor effect.
For progressively smaller NS angular velocities, $\Omega_*$, the
transition layer between the self-similar boundary layer solution
and the hot settling flow, $A\la D\la R_*$, becomes more and more extended,
liniting the applicability of our self-similar solution. Moreover, 
for small $\Omega_*$, the expression for $\dot J$ used above becomes 
inaccurate, too.Hence we limit our investication to the case of a 
rapidly rotating neutron star. 

A similar analysis can be applied to equation (\ref{momnt}). 
However, it is sufficient to prove that the omitted terms
are sub-dominant. Then the first order corrections to the
self-similar solution will also be small. Firstly, we readily see that
the second term in the first brackets is small:
$v^2/c_s^2\propto d^{2/5}\to 0$ as $d\to 0$.
Secondly, for the second term in the second brackets be sub-dominant,
$c_s^2$ should scale slower then linearly in $d$, i.e.,
if $c_s^2\propto d^\beta$ then $\beta$ must be less then unity.
From our solutions (\ref{sol2T}) and (\ref{sol1T}), we have  $\beta=2/5<1$.
We should comment here that with $d\Omega/dD\simeq0$ viscous heating
is negligible in the boundary layer. From equation (\ref{45r}) it is clear
that it is adiabatic compression of gas, $d\rho/dD>0$ that keeps the gas
hot in the presence of Bremsstrahlung cooling.

\begin{figure}
\psfig{file=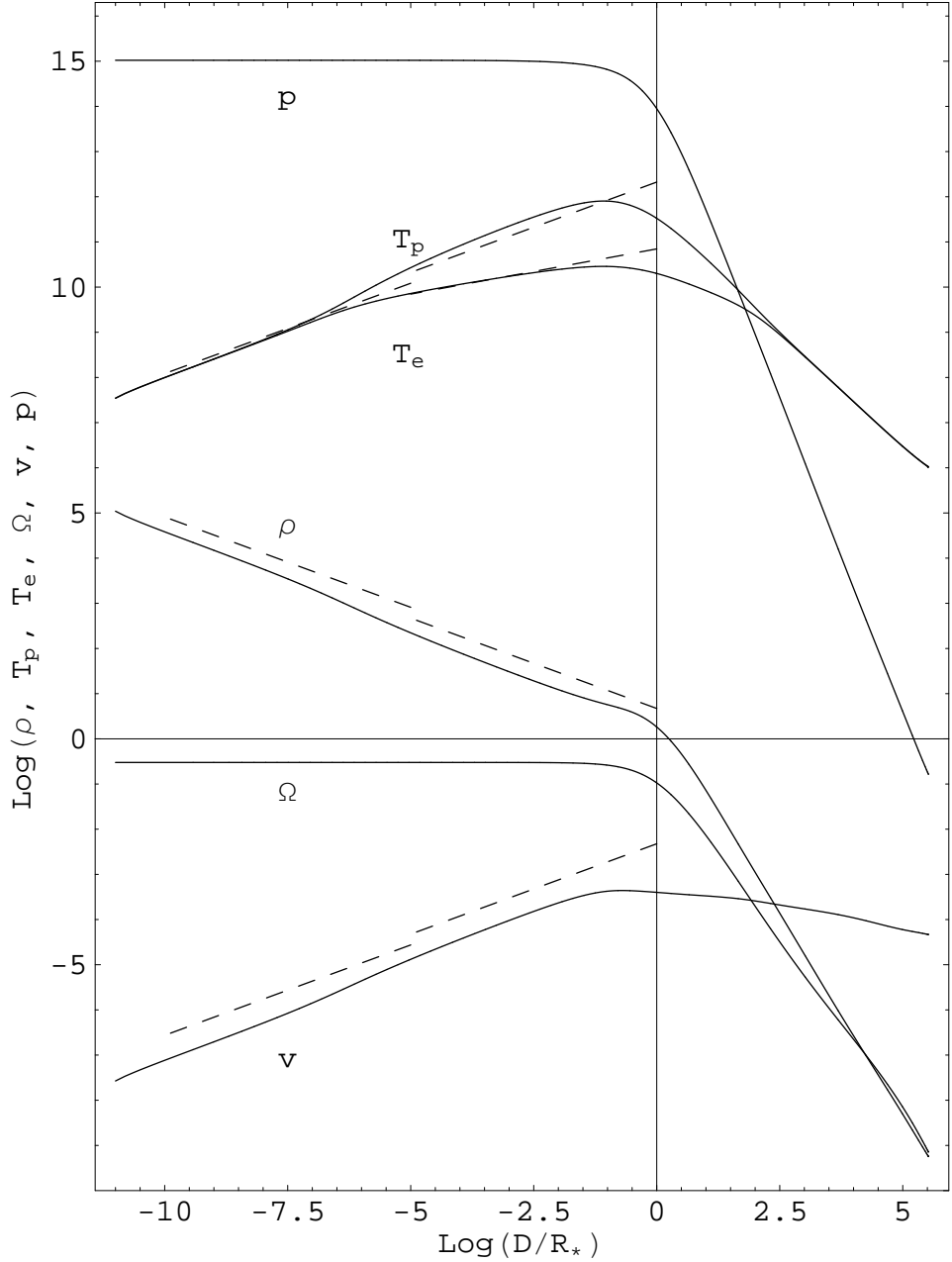,width=5in}
\caption{The structure of the boundary layer. $Log$ of $\rho$ 
(in $10^{18}$~cm$^{-3}$), $T_p,\ T_e$ (in Kelvin),
$\Omega/\Omega_{K*},\ v/c$~ and $p=\rho c_s^2$ (in dyne~cm$^{-2}$)
as functions of $Log(D/R_*)$ are shown
for $\dot m=0.01,\ \alpha=0.1,\ s=0.3$.
Grayscale lines represent the self-similar solution.The temperatures 
and the velocity are sensitive to the value of the external pressure,
which is estimated from the hot settling flow self-similar solution
\citep{MN01}. The agreenemt becomes nearly perfect when $p_{\rm out}$
is increased by a factor of 4, compared to Eq. (\ref{pout}).
\label{f:BL} }
\end{figure}

Finally, we confirm our theoretical results numerically. We use
the numerical code used in our previous studies, which solves
the system of the height-integrated viscous, hydrodynamic equations
using the relaxation method. The inhomogeneous grid was specially 
designed so that to resolve the thin boundary layer accurately.
For more discussion, we refer the reader to our previous paper
\citep{MN01}. The calculated structure of the boundary 
layer is presented in Figure \ref{f:BL}.
Note the remarkable agreement of this numerical solution with the 
theoretical one: $\Omega=$const., $p\approx$const., and 
$\rho,\ T_p,\ T_e,\ v$~ following the predicted scalings.

\section{Discussion}
\label{s:discuss}

It is instructive to summarize the scalings of various gas properties
as functions of stellar and accretion parameters. 
\bea
\rho&\propto&\alpha^{1/5}s^{2/5}\dot m^{2/5} m^{-1}r_*^{-9/5}d^{-2/5},\\
T_p&\propto&\alpha^{4/5}s^{8/5}\dot m^{-2/5} m^0 r_*^{-6/5}d^{2/5},\\
v&\propto&\alpha^{-1/5}s^{-2/5}\dot m^{3/5} m^0 r_*^{-1/5}d^{2/5}.
\eea
Here we remind that $r_*=R_*/R_g$ whereas $d=D/R_*$ and we used that
$T_p\propto c_{sp}^2$.

The luminosity per logarithmic interval in $D$, 
i.e., $dL\sim q^- R_*^2dD\sim q^-R_*^2 D\, d(\log{D})$, is then
\beq
dL/d(\log{D})\propto\alpha^{4/5}s^{8/5}\dot m^{3/5}m^{1}r_*^{-6/5}d^{2/5}.
\eeq 
Hence, most of the energy is radiated far away from the NS surface.
One may also note the strong dependence of $dL/d(\log{D})$ on the 
NS spin and radius. The transition radius, Eq. (\ref{d-tr}), scales as
\beq
d_{\rm tr}\simeq(6r_*c_{sp0}^2/c^2)^{-5/2}
\propto\alpha^{-4/5}s^{-8/5}\dot m^{2/5}m^0r_*^{1/5}.
\eeq
The total boundary layer luminosity is then estimated to be:
\beq
L_{BL}\propto\alpha^{4/5}s^{8/5}\dot m^{3/5}m^{1}r_*^{-6/5}d^{2/5}_{\rm tr}
\propto\dot m m r_*^{-1},
\eeq
which perfectly agrees with a simple energetic argument that
$L_{BL}\sim\dot Mc^2/r\sim \dot M_{\rm Edd}\dot m/r_*\sim m \dot m r_*^{-1}$
\citep{MN01}.

Our self-similar  solution is subsonic (otherwise a shock would form) 
with the Mach number of the infalling gas being
\beq
M_{}=v/c_s\propto d^{1/5},
\eeq
that is, if the gas is subsonic where the boundary layer matches the
bulk accretion flow (which is {\em always} so, because the gas in the
hot settling flow is highly subsonic), it will remain subsonic all the
way down to the stellar surface.

Next, we estimate the effect of Comptonization. The electron scattering
optical depth is $\tau_{\rm es}\simeq\rho\kappa_{\rm es}D$, where 
$\kappa_{\rm es}=\sigma_T/m_p$ is the electron scattering opacity 
for ionized hydrogen.
We can use our self-solution for $\rho$ to calculate $\tau_{\rm es}$.
Alternatively, we may recall that the density matches that of the
hot settling flow at the transition distance, and at in this region
the electron scattering optical depth \citep{MN01}, 
$\tau_{\rm es}\simeq 10^3\alpha s^2 r_*^{-1}$ (assuming again that
$D_{\rm tr}\ll R_*$), is $\lesssim1/3$ for typical parameters
$\alpha=0.1$,~ $s=0.1$ and $r_*=3$. Using the scaling for the density, 
$\rho\propto d^{-2/5}$, we estimate the optical depth in the
boundary layer as
\beq
\tau_{\rm es,BL}\sim10^3\alpha s^2 r_*^{-1}(d/d_{\rm tr})^{3/5}
\sim(1/3)\alpha_{0.1} s^2_{0.1} r_{*,3}^{-1}(d/d_{\rm tr})^{3/5}<1,
\eeq
where $\alpha_{0.1}=\alpha/0.1$ and similarly for other parameters.
Hence, gas is optically thin to electron scattering for 
reasonable accretion parameters. The Compton $y$-parameter, 
$y=16(c_{se}/c)^4\tau_{\rm es}$, calculated for the boundary layer 
should match that of the hot settling flow at the transition distance.
Using the result of \citet{MN01} and that $c_{se}^2\propto d^{1/5}$, 
we estimate
\beq
y\simeq 2\times10^6\alpha s^2r_*^{-2}(d/d_{\rm tr})
\sim 220\alpha_{0.1} s^2_{0.1}r_{*,3}^{-2}(d/d_{\rm tr}).
\eeq
The effect of Comptonization is important in the high-temperature
region of the boundary layer where $y\ga1$, which occurs at distances
larger or comparable to 
\beq
d_c\sim4.5\times10^{-3}\alpha_{0.1}^{-1} s^{-2}_{0.1}r_{*,3}^{2}d_{\rm tr}.
\eeq
Therefore, we conclude that Comptonization is not important deep inside
the boundary layer, $d<d_c$, and our one-temperature self-similar solution 
is accurate there. However, the electron temperature profile in the 
two-temperature zone may be substantially modified by the inverse 
Compton scattering. Since, however, the electron-proton collisions 
are relatively rare (the plasma is two-temperature), other gas
properties, such as the density, proton temperature, etc. should not 
be strongly affected. (Since the Coulomb collision rate increases with the 
decresing electron temperature, a more careful analysis is necessary. 
Such an analysis is beyond the scope this paper.) Note that in systems 
with slow rotation  ($s<0.01$) and with low-viscosity gas ($\alpha<0.01$), 
the structure of the boundary layer is not affected by Comptonization because
$d_c\ga d_{\rm tr}$.

Finally, we estimate the position of the photosphere, i.e., 
the distance at which the free-free optical depth becomes
of order unity. The free-free opacity is approximately equal to
$\alpha_{\rm ff}\simeq 1.7\times 10^{-25} T^{-7/2} (\rho/m_p)^2$~cm$^{-1}$.
Using our self-similar solution, we calculate the free-free optical
depth:
\beq
\tau_{\rm ff}=5.0\times10^{-25}\alpha^{-12/5} s^{-24/5} \dot m^{11/5}
m^{-1}r_*^{8/5} d^{-6/5}.
\eeq
The optical depth becomes greater than unity at distances below
the photospheric radius:
\beq
d_{\rm phot}=5.2\times10^{-18}\alpha^{-2}_{0.1} s^{-4}_{0.1} 
\dot m^{11/6}_{0.01} m^{5/6} r_{*,3}^{4/3}\ll 1.
\eeq
Hence, the free-free emission in, essentially, the entire boundary 
layer is optically thin.

\section{Conclusions}

In this paper we presented the analytical self-similar solution 
describing the boundary layer, which forms in the vicinity of a 
spinning neutron star. Our solution is hot, highly subsonic and 
contains no shocks. The crucial difference of our work from 
others is that the bulk accretion flow is a quasi-spherical
hot settling accretion flow, rather than a thin Shakura-Sunyaev disk.
We believe, our solution may be relevant to a boundary layer of an 
advection-dominated accretion flow as well. We critically examined 
the limitations of our solution, especially the effect of Comptonization.
We concluded that the one-temperature solution is accurate,
whereas the two-temperature solution accurately represents the
density, proton temperature, infall and angular velocity profiles, 
but the electron temperature profile may be modified by Comptonization.

\acknowledgements

The author is grateful to Ramesh Narayan and an anonymous referee
for insightful comments on the manuscript. This work has been 
supported by NASA and DoE.

\end{document}